\documentclass[12pt]{article}
\usepackage{graphicx}
\let\Large=\large
\newcommand{\alt}{\mathrel{\raisebox{-.6ex}{$\stackrel{\textstyle<}{\sim}$}}}
\newcommand{\agt}{\mathrel{\raisebox{-.6ex}{$\stackrel{\textstyle>}{\sim}$}}}

\begin{document}
\centerline{                    \hfill           {\bf UIC-HEP/97-4}}
\centerline{                    \hfill        {\bf UTEXAS-HEP-97-7}}
\centerline{                    \hfill        {\bf DOE-ER40757-098}}
\centerline{                    \hfill        {\bf UCD-HEP-98-30}}
\vspace*{.5in}

\centerline{\Large  Phase Effect of A General Two-Higgs-Doublet Model}
\centerline{\Large  in  $b\to s\gamma$}
\vskip 1cm

\begin{center}
David Bowser-Chao$^{a)}$, Kingman Cheung$^{c,b)}$, and Wai-Yee Keung$^{a)}$\\
\it
$^{a)}$Department of Physics, University of Illinois at Chicago, Chicago,
IL 60625 \\
$^{b)}$Center for Particle Physics, University of Texas, Austin, TX 78712\\
$^{c)}$Department of Physics, University of California, Davis, CA 95616\\
\end{center}

\begin{abstract}
In a general two-Higgs-doublet model (2HDM),
without the {\it ad hoc} discrete symmetries to prevent tree-level
flavor-changing-neutral currents, an extra phase angle in the
charged-Higgs-fermion coupling is allowed.
We show that the charged-Higgs amplitude interferes destructively or
constructively with the standard model amplitude depending crucially on
this phase angle.
The popular model I and II are special cases of our analysis.
As a result of this phase angle the severe constraint on
the charged-Higgs boson mass imposed by the inclusive rate of $b\to s\gamma$
from CLEO can be relaxed.
We also examine the effects of this phase angle on the neutron electric
dipole moment.  Furthermore, we also discuss other constraints on the
charged-Higgs-fermion couplings coming from measurements of
$B^0-\overline{B^0}$ mixing, $\rho_0$, and $R_b$.
\end{abstract}

\thispagestyle{empty}

\section{Introduction}

One of the most popular extensions of the standard model (SM) is the
two-Higgs-doublet model (2HDM) \cite{hunter}, which has two complex
Higgs doublets instead of only one in the SM.  The 2HDM allows
flavor-changing neutral currents (FCNC),
which can be avoided by imposing an {\it ad hoc} discrete symmetry \cite{glas}.
One possibility to avoid the FCNC is to couple the fermions only to
one of the two
Higgs doublet, which is often known as model I.  Another possibility is to
couple the first Higgs doublet to the down-type quarks while the second
Higgs doublet to the up-type quarks, which is known as model II.  This model II
has been very popular because it is the building block of the minimal
supersymmetric standard model.  The physical content of the Higgs sector
includes
a pair of CP-even neutral Higgs bosons $H^0$ and $h^0$ , a CP-odd neutral
boson $A$, and a pair of charged-Higgs bosons $H^\pm$.

Models I and II have been extensively studied in literature and tested
experimentally.  One of the most stringent tests is the radiative decay of
$B$ mesons, specifically, the inclusive decay rate of $b\to s \gamma$,
which has the least hadronic uncertainties.
The SM rate of $b\to s\gamma$ including the
improved leading-order logarithmic QCD corrections is
predicted \cite{buras} to be $(2.8\pm 0.8) \times 10^{-4}$, of
which the uncertainty mainly comes from the factorization scale and from
the next-to-leading order corrections.
\footnote{The NLO order calculations for the SM and 2HDM I and II are
available very recently \cite{nlo}. The SM result is $(3.29\pm 0.33)\times
10^{-4}$, which is consistent with the LO calculation.  However, the
NLO calculation is not available for 2HDM III and, therefore, we will use
the LO result consistently throughout the paper.}
In the 2HDM,
the rate of $b\to s\gamma$ can be enhanced substantially for large regions
in the parameter space of the mass $M_{H^\pm}$ of
the charged-Higgs boson and $\tan \beta=v_2/v_1$,  where $v_1$ and
$v_2$ are the vacuum expectation values of the two Higgs doublets.
CLEO published a result of $b\to
s\gamma$ inclusive rate of $(2.32\pm 0.57 \pm 0.35) \times
10^{-4}$ in 1995 \cite{cleo95},
which is recently updated to $(3.15\pm0.35\,({\rm stat})\,
\pm 0.32\,({\rm sys})\, \pm 0.26\, ({\rm mod}))\times 10^{-4}$ in 1998
\cite{cleo98}.
ALEPH also published a result of $(3.11\pm0.80\,({\rm stat})\,
\pm 0.72\,({\rm sys})\, )\times 10^{-4}$ \cite{aleph}.
The 95\%CL limit published by CLEO is also updated to
$2\times 10^{-4}< B(b\to s \gamma)< 4.5\times 10^{-4}$ \cite{cleo98}.
The data is now more consistent with the SM prediction than before.
Hence, the
experimental result puts a rather stringent constraint on the 
charged-Higgs boson mass $M_{H^\pm}$ and $\tan \beta$.
In model II, the constraint is $M_{H^\pm}\agt 350$ GeV for $\tan\beta$ larger
than 1, and even stronger for smaller $\tan\beta$ \cite{hewett}.

Recently, there have been some studies \cite{recent,atwood}
on a more general 2HDM without the discrete symmetries as in
models I and II.  It is often referred as model III.  FCNC's in general exist
in model III.  However, the FCNC's involving the first two
generations are highly suppressed from low-energy experiments, and those
involving
the third generation is not as severely suppressed as the first two
generations.  It implies that model III should be parameterized in a way to
suppress the tree-level FCNC couplings of the first two generations while
the tree-level FCNC couplings involving the third generation can be made
nonzero as long as they do not violate any existing experimental data,
e.g., $B^0-\overline{B^0}$ mixing.

In this work, we simply assume all tree-level FCNC couplings
to be negligible.
Even though in such a simple model the
couplings involving Higgs bosons and fermions can have complex phases
$e^{i\theta}$.
The effects of such extra phases in $b\rightarrow s\gamma$
have been noticed in Ref.\cite{Wolfenstein}.
%
%
In this paper, we shall study carefully the constraint on the phase angle
in the product, $\lambda_{tt}\lambda_{bb}$, of Higgs-fermion couplings
(see below)
versus the mass of the charged-Higgs boson from the CLEO data of
$b\to s\gamma$.  We shall show that in the calculation of $b\to s\gamma$ the
charged-Higgs amplitude interferes destructively or constructively
with the SM amplitude depending crucially on this phase angle and less
on the charged-Higgs mass.  The usual model I and II are special cases
in our study.  We shall also
show that the previous constraints on the charged-Higgs mass and
$\tan\beta$ imposed by the CLEO data can be relaxed because of the
presence of this extra phase angle.
There are other processes in which the effects of the phase angle can be seen.
One of these that we study in this report is
the neutron electric dipole moment.
In addition, we also discuss the constraints from experimental measurements
of $B^0-\overline{B^0}$ mixing, $\rho_0$, and $R_b$.

The organization is as follows.  In the next section we describe the
content of the general 2HDM and write down the Feynman rules for model
III.  In Sec. III, we describe briefly the effective hamiltonian
formulation for the decay of $b\to s\gamma$ and derive the Wilson
coefficients in model III.  We present our numerical results for $b\to s\gamma$
and study the case of neutron electric dipole moment in Sec. IV.
In Sec V, we discuss other experimental constraints from
measurements of $B^0-\overline{B^0}$ mixing, $\rho_0$, and $R_b$.
Finally, we conclude in Sec. VI.

\section{The General Two-Higgs-Doublet model}

In a general two-Higgs-doublet model, both the doublets can couple to the
up-type and down-type quarks.  Without loss of generosity, we work in a basis
such that the first doublet generates all the gauge-boson and fermion
masses:
\begin{equation}
\langle \phi_1 \rangle = \left( \begin{array}{c} 0 \\
                               \frac{v}{\sqrt{2}} \end{array} \right )
\;, \qquad
\langle \phi_2 \rangle = 0
\end{equation}
where $v$ is related to the $W$ mass by $M_W={g\over2}v$.
In this basis, the first doublet $\phi_1$ is the same as the SM doublet, while
all the new Higgs fields come from the second doublet $\phi_2$.  They are
written as
\begin{equation}
\phi_1=\frac{1}{\sqrt 2} \left( \begin{array}{c}
                             \sqrt{2} G^+ \\
                             v + \chi_1^0 + i G^0 \end{array} \right )
\; , \qquad
\phi_2=\frac{1}{\sqrt 2} \left( \begin{array}{c}
                               \sqrt{2} H^+ \\
                               \chi_2^0 + i A^0 \end{array} \right ) \;,
\end{equation}
where $G^0$ and $G^\pm$ are the Goldstone bosons that would be eaten away
in the Higgs mechanism to become the longitudinal components of the weak
gauge bosons.  The $H^\pm$ are the physical charged-Higgs bosons and $A^0$ is
the physical CP-odd neutral Higgs boson.  The $\chi_1^0$ and $\chi_2^0$ are
not physical mass eigenstates but linear combinations of the CP-even
neutral Higgs bosons:
\begin{eqnarray}
\chi_1^0 &=& H^0 \cos\alpha  - h^0 \sin \alpha  \\
\chi_2^0 &=& H^0\sin\alpha  + h^0 \cos \alpha  \;,
\end{eqnarray}
where $\alpha$ is the mixing angle.  In this basis, there is no couplings
of $\chi_2^0 ZZ$ and $\chi_2^0 W^+ W^-$.
We can write down\cite{atwood} the Yukawa Lagrangian for model III as
\begin{equation}
-{\cal L}_Y = \eta_{ij}^U \overline{Q_{iL}} \tilde{\phi_1} U_{jR} +
              \eta_{ij}^D \overline{Q_{iL}} \phi_1 D_{jR} +
       \xi_{ij}^U \overline{Q_{iL}} \tilde{\phi_2} U_{jR} +
       \xi_{ij}^D \overline{Q_{iL}} \phi_2 D_{jR}   \quad \; + {\rm h.c.} \;,
\end{equation}
where $i,j$ are generation indices, $\tilde{\phi}_{1,2} = i\sigma_2 
\phi_{1,2} \;$, $\eta_{ij}^{U,D}$ and $\xi_{ij}^{U,D}$
are, in general, nondiagonal coupling matrices, and $Q_{iL}$ is the
left-handed fermion doublet and $U_{jR}$ and $D_{jR}$ are the right-handed
singlets.  Note that these $Q_{iL}$, $U_{jR}$, and $D_{jR}$ are weak
eigenstates, which can be rotated into mass eigenstates.
As we have mentioned above, $\phi_1$ generates all the
fermion masses and, therefore, $\frac{v}{\sqrt{2}}\eta^{U,D}$ will become
the up- and down-type quark-mass matrices after a bi-unitary transformation.
After the transformation the Yukawa Lagrangian becomes
\begin{eqnarray}
{\cal L}_Y &=& -\overline{U} M_U U - \overline{D} M_D D
   - \frac{g}{2M_W} (H^0\cos\alpha - h^0 \sin\alpha)
     \biggr(\overline{U} M_U U + \overline{D} M_D D \biggr ) \nonumber \\
&+& \frac{ig}{2 M_W} G^0 \left(\overline{U} M_U \gamma^5 U
                             - \overline{D} M_D \gamma^5 D \right)\nonumber \\
&+& \frac{g}{\sqrt{2}M_W} G^- \overline{D} V^\dagger_{\rm CKM} \biggr [
     M_U \hbox{$1\over2$}(1+\gamma^5) - M_D \hbox{$1\over2$}(1-\gamma^5) \biggr]
 U \nonumber \\
&-& \frac{g}{\sqrt{2}M_W} G^+ \overline{U} V_{\rm CKM} \biggr [
     M_D \hbox{$1\over2$}(1+\gamma^5) - M_U \hbox{$1\over2$}(1-\gamma^5) \biggr]
 D  \nonumber \\
&-&
\frac{H^0 \sin\alpha + h^0\cos\alpha }{\sqrt{2}} \Biggr[
  \overline{U} \biggr( {\hat\xi}^U \hbox{$1\over2$}(1+\gamma^5) + {\hat\xi}^{U\dagger}
 \hbox{$1\over2$}(1-\gamma^5)
         \biggr ) U  \nonumber \\
&& +\overline{D} \biggr( {\hat\xi}^D \hbox{$1\over2$}(1+\gamma^5) +
 {\hat\xi}^{D\dagger} \hbox{$1\over2$}(1-\gamma^5)
         \biggr ) D
  \Biggr ]  \nonumber \\
&+& \frac{i A^0}{\sqrt{2}} \Biggr [
    \overline{U} \biggr( {\hat\xi}^U
           \hbox{$1\over2$}(1+\gamma^5) -{\hat\xi}^{U\dagger}
           \hbox{$1\over2$}(1-\gamma^5)
        \biggr ) U
   -\overline{D} \biggr( {\hat\xi}^D
           \hbox{$1\over2$}(1+\gamma^5) - {\hat\xi}^{D\dagger}
           \hbox{$1\over2$}(1-\gamma^5)
        \biggr ) D
     \Biggr ] \nonumber \\
&-&  H^+ \overline{U} \biggr[ V_{\rm CKM} {\hat \xi}^D
             \hbox{$1\over2$}(1+\gamma^5) -
   {\hat\xi}^{U\dagger} V_{\rm CKM}
             \hbox{$1\over2$}(1-\gamma^5) \biggr] D
  \nonumber \\
&-&  H^- \overline{D} \biggr[ {\hat \xi}^{D\dagger} V^\dagger_{\rm CKM}
   \hbox{$1\over2$}(1-\gamma^5) -
      V^\dagger_{\rm CKM} {\hat\xi}^U \hbox{$1\over2$}(1+\gamma^5) \biggr] U
 \;\;,
\label{rule}
\end{eqnarray}
where
$U$ represents the mass eigenstates of $u,c,t$ quarks and
$D$ represents the mass eigenstates of $d,s,b$ quarks.
The transformations are defined by
$M_{U,D}=\hbox{diag}(m_{u,d}, m_{c,s}, m_{t,b})= {v\over\sqrt{2}}
({\cal L}_{U,D})^{\dagger} \eta^{U,D} ({\cal R}_{U,D})$,
$\hat\xi^{U,D} = ({\cal L}_{U,D})^{\dagger} \xi^{U,D} ({\cal R}_{U,D})$.
The Cabibbo-Kobayashi-Maskawa matrix \cite{ckm} is
$V_{\rm CKM}=({\cal L}_U)^\dagger({\cal L}_D)  $.

The FCNC couplings are contained in the matrices $\hat\xi^{U,D}$.  A
simple ansatz for $\hat\xi^{U,D}$ would be \cite{recent}
\begin{equation}
\label{anat}
\hat\xi^{U,D}_{ij} = \lambda_{ij} \frac{g\sqrt{m_i m_j}}{\sqrt{2}M_W}
\end{equation}
by which the quark-mass hierarchy ensures that the FCNC within the
first two generations are naturally suppressed by the small quark
masses, while a larger freedom is allowed for the FCNC involving the
third generations.  Here $\lambda_{ij}$'s are of order $O(1)$ and unlike
previous studies \cite{recent,atwood} they can be complex, which
give nontrivial consequences
different from previous analyses based on model I and II.
An interesting example would be the inclusive rate of $b\to s\gamma$
that we shall study next.  Such complex $\lambda_{ij}$'s allow the
charged-Higgs amplitude to interfere destructively or constructively
with the SM amplitude.
As we have mentioned, models I and II are special cases in our
study and so the previous constraints\cite{hewett} imposed on the
charged-Higgs mass and $\tan\beta$ by the CLEO data can be relaxed by
the presence of the extra phase angle.
Other interesting phenomenology of the complex $\lambda_{ij}$'s
includes the electric dipole moments of electrons and
quarks\cite{dave} as a consequence of the explicit CP violation due to
the complex phase in the charged-Higgs sector.
For simplicity we choose
$\hat\xi^{U,D}$ to be diagonal to suppress all tree-level FCNC couplings
 and, consequently, the $\lambda_{ij}$'s
are also diagonal but remain complex.  Such a simple scenario is
sufficient to demonstrate our claims.

\section{Inclusive $B\to X_s \gamma$}

The detail description of the effective hamiltonian approach can be found in
Refs. \cite{buras,wise}.  Here we present the highlights that are relevant
to our discussions.  The effective hamiltonian for $B\to X_s\gamma$ at a
factorization scale of order $O(m_b)$ is given by
\begin{equation}
\label{eff}
{\cal H}_{\rm eff} = - \frac{G_F}{\sqrt{2}} V^*_{ts} V_{tb} \biggr [
\sum_{i=1}^6 \; C_i(\mu) Q_i(\mu) + C_{7\gamma}(\mu) Q_{7\gamma}(\mu) +
C_{8G}(\mu) Q_{8G}(\mu) \biggr ] \;.
\end{equation}
The operators $Q_i$ can be found in Ref.\cite{buras}, of which the $Q_1$ and
$Q_2$ are the current-current operators and $Q_3-Q_6$ are QCD penguin
operators.  $Q_{7\gamma}$ and $Q_{8G}$ are, respectively, the magnetic
penguin operators specific for $b\to s\gamma$ and $b\to s g$.  Here we also
neglect the mass of the external strange quark compared to the external
bottom-quark mass.

The factorization in Eq.(\ref{eff}) facilitates the separation of the
short-distance and long-distance parts, of which the short-distance parts
correspond to the Wilson coefficients $C_i$ and are calculable
by perturbation while the long-distance parts correspond to the operator
matrix elements.
The physical quantities based on Eq.~(\ref{eff}) should
be independent of the factorization scale $\mu$.  The natural scale for
factorization is of order $m_b$ for the decay $B\to X_s \gamma$.
The calculation of the $C_i(\mu)$'s divides into two separate steps.
First, at the electroweak scale, say $M_W$, the full theory is matched onto
the effective theory and the coefficients $C_i(M_W)$ at the $W$-mass scale
are extracted in the matching process.
In a while, we shall present these coefficients $C_i(M_W)$ in our model III.
Second, the coefficients $C_i(M_W)$ at the $W$-mass scale are evolved
down to the bottom-mass scale using renormalization group equations.
Since the operators $Q_i$'s are all mixed under renormalization, the
renormalization group equations for $C_i$'s are a set of coupled
equations:
\begin{equation}
\vec C(\mu) = U(\mu,M_W) \vec C(M_W) \;,
\end{equation}
where $U(\mu,M_W)$ is the evolution matrix and $\vec C(\mu)$ is the vector
consisting of $C_i(\mu)$'s.  The calculation of the entries of the evolution
matrix $U$ is nontrivial but it has been written down completely in the
leading order \cite{buras}.  The coefficients $C_i(\mu)$ at the scale $O(m_b)$
are given by \cite{buras}
\begin{eqnarray}
C_j(\mu) &=& \sum_{i=1}^8 k_{ji} \eta^{a_i} \qquad (j=1, ..., 6) \;\;,\\
\label{c7}
C_{7\gamma}(\mu) &=& \eta^{\frac{16}{23}} C_{7\gamma}(M_W) + \hbox{$8\over3$}
\left(\eta^{\frac{14}{23}} - \eta^{\frac{16}{23}} \right ) C_{8G}(M_W)
+ C_2(M_W) \sum_{i=1}^8 h_i \eta^{a_i} \;\;,\\
C_{8G}(\mu) &=& \eta^{\frac{14}{23}} C_{8G}(M_W) + C_2(M_W) \sum_{i=1}^8
\bar h_i \eta^{a_i} \;\;,
\end{eqnarray}
with $\eta=\alpha_s(M_W)/\alpha_s(\mu)$.  The $a_i$'s, $k_{ji}$'s,
$h_i$'s, and $\bar h_i$'s can be found in Ref.~\cite{buras}.

Once we have all the Wilson coefficients at the scale $O(m_b)$ we can then
compute the decay rate of $B\to X_s \gamma$.  The decay amplitude for
$B\to X_s \gamma$ is given by
\begin{equation}
{\cal A}(B\to X_s\gamma) = -\frac{G_F}{\sqrt{2}} V^*_{ts} V_{tb}
C_{7\gamma}(\mu) \langle Q_{7\gamma} \rangle \;,
\end{equation}
in which we use the spectator approximation to evaluate the matrix element
$\langle Q_{7\gamma} \rangle$ and $m_B \simeq m_b$.  The decay rate
of $B\to X_s\gamma$ is given by
\begin{equation}
\Gamma(B\to X_s\gamma) = \frac{G_F^2 |V^*_{ts} V_{tb}|^2 \alpha_{\rm em} m_b^5}
{32\pi^4} \; |C_{7\gamma}(m_b)|^2 \;,
\end{equation}
where $C_{7\gamma}(m_b)$ is given in Eq.~(\ref{c7}).
Since this decay rate depends on the fifth power of $m_b$, a small uncertainty
in the choice of $m_b$ will create a large uncertainty in the decay rate,
therefore, the decay rate of $B\to X_s \gamma$ is often normalized to the
experimental semileptonic decay rate as
\begin{equation}
\label{cal}
\frac{\Gamma(B\to X_s\gamma)}{\Gamma(B\to X_c e \bar \nu_e)} =
\frac{|V^*_{ts} V_{tb}|^2}{|V_{cb}|^2} \, \frac{6\alpha_{\rm em}}
 {\pi f(m_c/m_b)} |C_{7\gamma}(m_b)|^2 \;,
\end{equation}
where $f(z) = 1-8z^2 +8z^6 - z^8 -24z^4 \ln z$.

The remaining task is the calculation of the Wilson coefficients $C_i(M_W)$ at
the $W$-mass scale.  The necessary Feynman rules can be obtained from the
Lagrangian in Eq.~(\ref{rule}).  As we have mentioned, we assume all
tree-level FCNC couplings negligible and, therefore, the neutral-Higgs bosons
do not contribute at tree level or at one-loop level.  The only contributions
at one-loop level come from the charged-Higgs bosons $H^\pm$, the charged
Goldstone bosons $G^\pm$, and the SM $W^\pm$ bosons.

The coefficients $C_i(M_W)$ at the leading order in model III
are given by
\begin{eqnarray}
C_j(M_W) &=& 0 \qquad (j=1,3,4,5,6) \;, \label {c6mw}\\
C_2(M_W) &=& 1 \;,\label{c2mw} \\
C_{7\gamma}(M_W) &=& - \frac{A(x_t)}{2} - \frac{A(y)}{6} |\lambda_{tt}|^2
                     +B(y) \lambda_{tt} \lambda_{bb}  \;,\label{c7mw} \\
C_{8G} (M_W) &=& -\frac{D(x_t)}{2} - \frac{D(y)}{6} |\lambda_{tt}|^2 +
                     E(y) \lambda_{tt}\lambda_{bb} \label{c8mw} \;\;,
\end{eqnarray}
where $x_t = m_t^2/M_W^2$, and $y = m_t^2 / M_{H^\pm}^2$.
The Inami-Lim functions\cite{japan} are given by
\begin{eqnarray}
A(x) &=& x \biggr [ \frac{8x^2 +5x -7}{12(x-1)^3} - \frac{(3x^2 -2x)\ln x}
                           {2(x-1)^4} \biggr ] \\
B(y)&=&y\biggr[\frac{5y-3}{12 (y-1)^2} - \frac{(3y-2) \ln y}{6(y-1)^3}\biggr ]
\\
D(x)&=& x\biggr[ \frac{x^2 - 5x -2}{4(x-1)^3} + \frac{3x\ln x}{2(x-1)^4}\biggr
] \\
E(y)&=& y\biggr[\frac{y-3}{4(y-1)^2} + \frac{\ln y}{2(y-1)^3} \biggr ] \;\;.
\end{eqnarray}
The SM results for the Wilson coefficients $C_i(M_W)$ for $i=1,...,6$
are the same as in Eqs.~(\ref{c6mw}) and (\ref{c2mw}),
while $C_{7\gamma}(M_W)$ and $C_{8G}(M_W)$ only have the first term as in
Eqs.~(\ref{c7mw}) and (\ref{c8mw}), respectively.
Thus, we already have all the necessary pieces to compute the decay rate
of $B\to X_s\gamma$.

Before we leave this section we would like to emphasize that the expressions
for $C_i(M_W)$ in Eqs.~(\ref{c6mw}) -- (\ref{c8mw}) obtained for model III
can be reduced to the results of models I and II by the following substitutions:
\begin{equation}
\lambda_{tt} \rightarrow \cot \beta \quad\hbox{and}
\qquad \lambda_{bb} \rightarrow \cot \beta \quad
\hbox{(for model I),}
\end{equation}
and
\begin{equation}
\lambda_{tt} \rightarrow  \cot \beta \quad\hbox{and}
\qquad \lambda_{bb} \rightarrow -\tan \beta\quad
\hbox{(for model II).}
\end{equation}

\section{Results}

We use the following inputs \cite{buras,langacker,pdg}
for our calculation: $m_t=173.8$ GeV, $M_W=80.388$ GeV,
$|V_{ts}^* V_{tb}|^2/|V_{cb}|^2 =0.95$,
$m_c/m_b=0.3$, and $B(b\to c e^- \bar \nu)=10.45\pm 0.21$\%,
$\alpha_{\rm em}(m_b)\simeq 1/133$, and
$\alpha_s(M_Z)=0.119$ and a 1-loop $\alpha_s$
is employed.  The branching ratio $B(B\to X_s \gamma)$ is calculated using
Eq.~(\ref{cal}).  The free parameters are then $M_{H^\pm}$, $\lambda_{tt}$,
and $\lambda_{bb}$, as in Eqs.~(\ref{c7mw}) and (\ref{c8mw}).

\begin{figure}[t]
\leavevmode
\begin{center}
\includegraphics[width=3.5in]{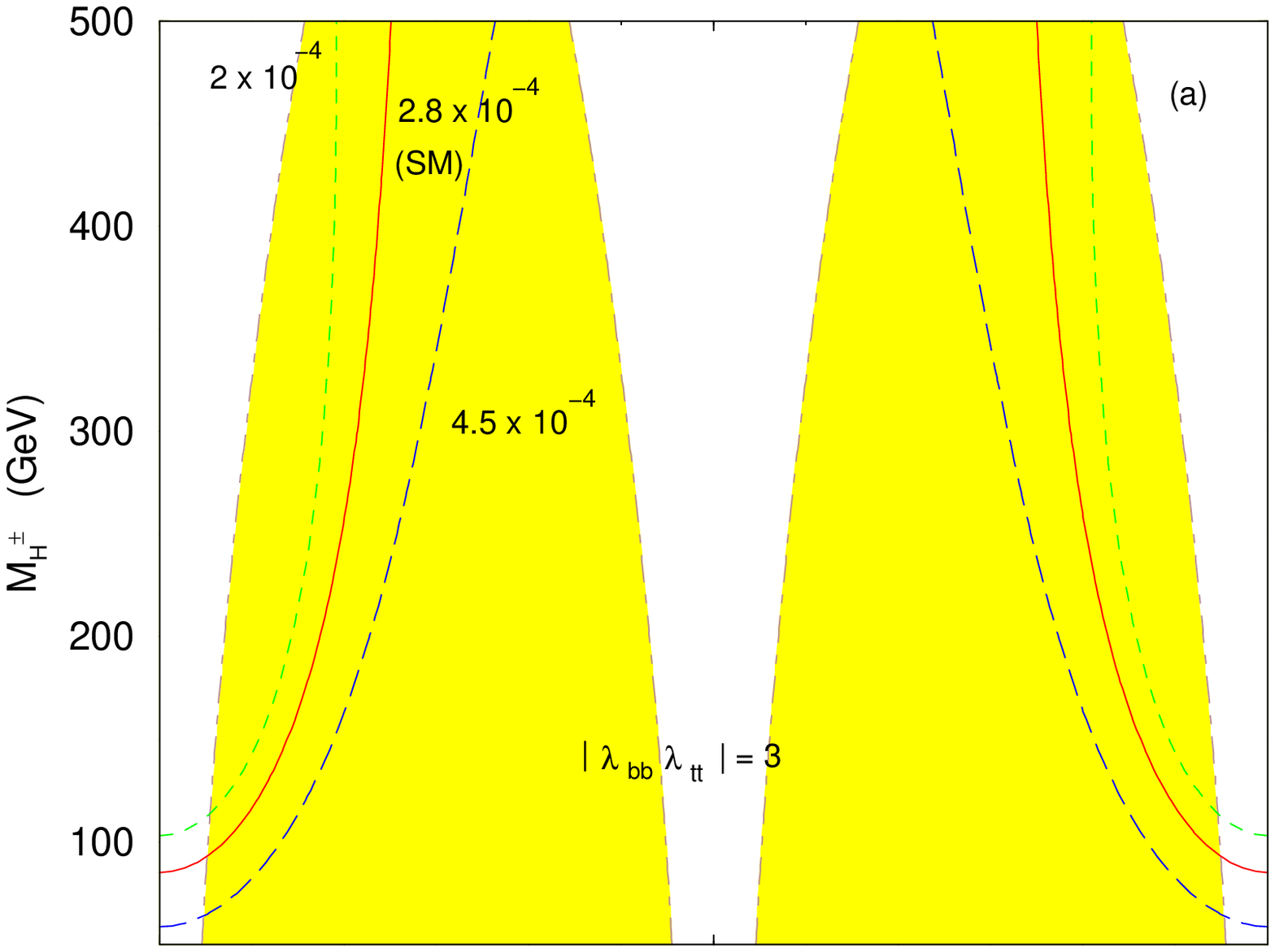}
\includegraphics[width=3.5in]{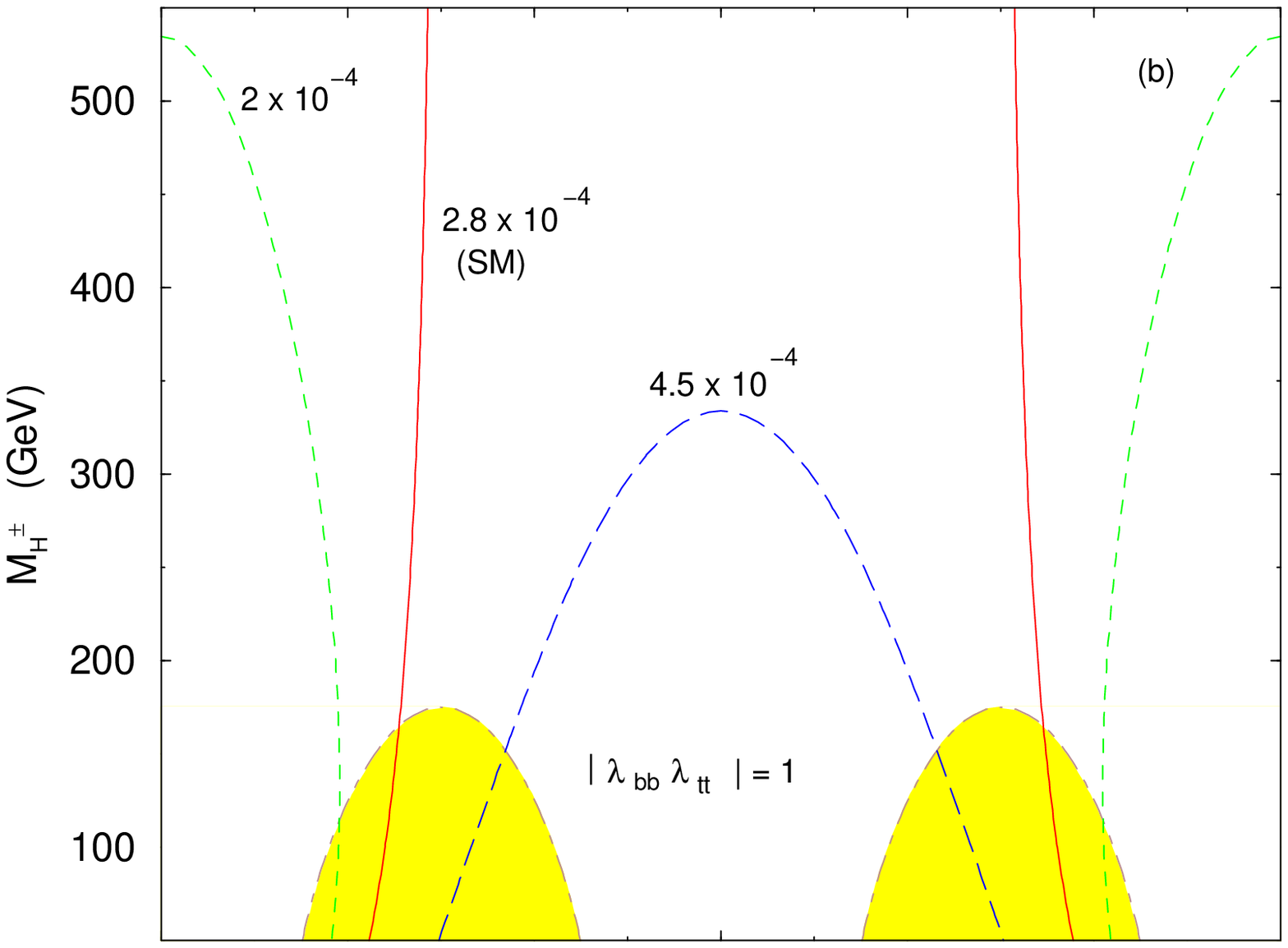}
\includegraphics[width=3.5in]{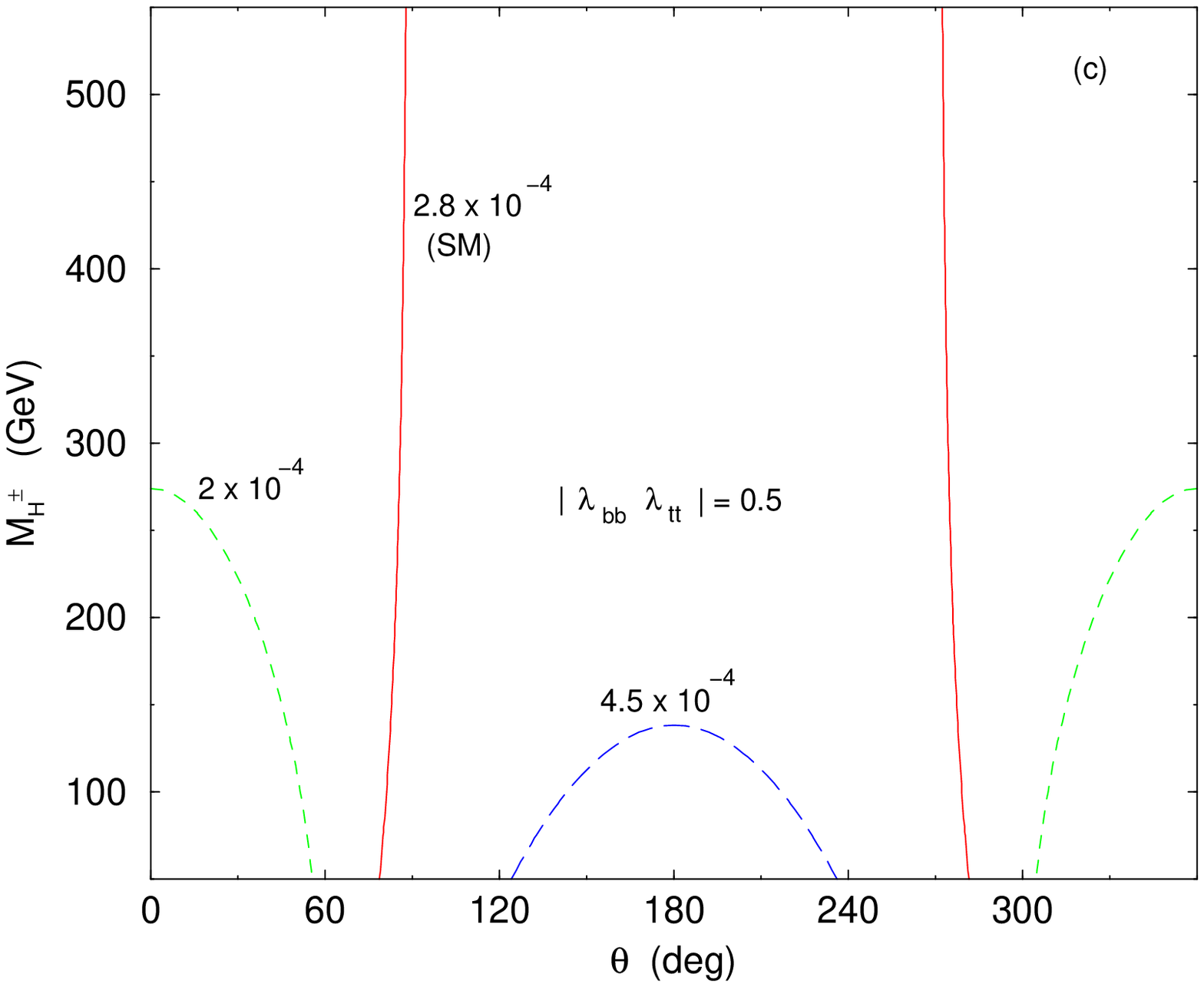}
\end{center}
\caption{Contour plot of the branching ratio $b\rightarrow s\gamma$
versus $M_H^\pm$ and the phase of $\lambda_{tt}\lambda_{bb}$ for
various values of $|\lambda_{tt}\lambda_{bb}|=3,1,0.5$.
The shaded areas are excluded by the NEDM constraint
$|d_n|<10^{-25}$ e$\cdot$cm.
}
\label{bs}
\end{figure}

Since the term proportional to $\lambda_{tt} \lambda_{bb}$ is, in general,
complex we let $\lambda_{tt} \lambda_{bb}=|\lambda_{tt} \lambda_{bb}|
e^{i\theta}$.  We show the contours of the branching ratio in the plane of
$\theta$ and $M_{H^\pm}$ for $|\lambda_{tt}\lambda_{bb}|=
3, 1, 0.5$ in Fig.~\ref{bs} (a), (b), and (c), respectively.
The contours are symmetric about $\theta=180^\circ$. 
The contours are $B=(2, 2.8, 4.5) \times 10^{-4}$,
which correspond to 95\%CL lower limit, the SM value, and the 95\%CL upper
limit.  The value of $|\lambda_{bb}|$ is set at 50 as
preferred in the $R_b$ constraint that will be shown in the
next section.  The corresponding values of $|\lambda_{tt}|$ are
$0.06, 0.02$, and $0.01$, which satisfy the constraint from the
$B^0-\overline{B^0}$ mixing, as will also be discussed in the next section.
Here the term proportional to $|\lambda_{tt}|^2$ is not crucial  because
the coefficient of $|\lambda_{tt}|^2$ is small compared with other two terms
in Eqs.~(\ref{c7mw}) and (\ref{c8mw}).

The results of the conventional model II (which can be obtained from our 
general results by the substitution: $\lambda_{tt} \to \cot \beta$, 
$\lambda_{bb} \to -\tan \beta$) can be read off from
Fig. \ref{bs}(b) at $\theta=180^\circ$.  The $b\to s\gamma$ data severely
constrains $M_{H^\pm}\agt 350$ GeV at 95\%CL level, because at
$\theta=180^\circ$
the SM amplitude interferes entirely constructively with the charged
Higgs-boson amplitude.  It is obvious
that at other angles the mass of
the charged Higgs-boson mass is less constrained, especially, in the range
$\theta=50^\circ-90^\circ$
the entire range of charged Higgs-boson mass is allowed
by the $b\to s\gamma$ constraint as long as $|\lambda_{tt} \lambda_{bb}|
\alt 1$.  However, when $|\lambda_{tt} \lambda_{bb}|$ is getting larger,
say 3,
(see Fig. \ref{bs}(a)) the allowed range of charged Higgs-boson mass becomes
narrow.  This is because the charged Higgs-boson amplitude becomes too large
compared with the SM amplitude.
On the other hand, when $|\lambda_{tt} \lambda_{bb}|$ becomes small
the allowed range
charged Higgs-boson mass is enlarged, as shown in Fig. \ref{bs}(c).
The significance of the phase angle $\theta$ is that the constraints previously
on $M_{H^\pm}$ and $\tan\beta$ are evolved into $\theta$, $M_{H^\pm}$,
$\lambda_{tt}$, and $\lambda_{bb}$, where we do not need to impose
$|\lambda_{tt}|=1/|\lambda_{bb}|$, as in model II.  The previous tight
constraint on $m_{H^\pm}$ is now relaxed down to virtually the direct search
limit of almost 60 GeV at LEPII \cite{ichep98}.

The phase $\theta$ of $\lambda_{tt}\lambda_{bb}$ can give rise to the
neutron electric dipole moment (NEDM). The physics involved can be
understood as follows. First, at the electroweak scale the phase
$\theta$ induces the CP violating color dipole moment (CDM) of the
$b$ quark. Second, the CDM of $b$ quark evolves by renormalization to
the scale at $m_b$ and turns into the Weinberg operator\cite{weinberg_CP}
({\it i.e.} the gluonic CDM\cite{braaten_dipole}) when the $b$-quark
field is integrated away. Finally, this gives NEDM at the nucleon mass
scale:
\begin{equation}
d_n^g =g_s^3(\mu) C_g(\mu) \langle {\cal O}_g(\mu) \rangle,
\quad\hbox{where }
{\cal O}_g=\hbox{$1\over6$}f^{abc}
\varepsilon^{\delta\nu\alpha\beta}
G^a_{\alpha\beta}G^b_{\lambda\delta}G^c_{\lambda\nu}
\ .
\end{equation}
Weinberg suggested the hadronic scale $\mu$ to be set at the value
such that $g_s(\mu)=4\pi/\sqrt{6}$. Instead we choose $\mu$ at the
nucleon mass.
The hadronic matrix element $\langle {\cal O}_g(\mu) \rangle$ is very
uncertain.  A typical estimate from the naive dimension analysis
(NDA)\cite{manohar} relates the matrix element to the chiral symmetry
breaking scale $M_\chi=2\pi F_\pi=1.19$ GeV,
\begin{equation}
{\cal O}_g= {eM_\chi\over 4\pi} \xi_g(\mu) \ .
\end{equation}
The parameter $\xi_g$ is set to be 1 in NDA, but other calculations
result in different $\xi_g$.
QCD sum rule performed by Chemtob\cite{chemtob} gives $\xi_g=0.07$.
Scaling argument by Bigi and Uraltsev\cite{bigi} yields a value
$\xi_g=0.03$.  We choose $\xi_g=0.1$ for our analysis. The Wilson
coefficient of the Weinberg operator $C_g$ evolves according to the RG
equation\cite{braaten} and matches\cite{boyd,ckly} that induced by the
CDM $C_b$ of the $b$ quark at the scale $m_b$. Our definitions of
Wilson coefficients follow the notation in Ref.\cite{ckly},
\begin{equation}
C_g(\mu)={1\over 32\pi^2} C_b(m_b)
\left( {\alpha_s(m_b)\over \alpha_s(m_c)} \right)^{54\over 25}
\left( {\alpha_s(m_c)\over \alpha_s(\mu)} \right)^{54\over 27}
\ .
\end{equation}
The CDM of the $b$ quark comes from the CP violation
of the charged Higgs coupling at the electroweak scale and at the scale
$m_b$ it is given by
\begin{equation}
C_b(m_b)
={\sqrt{2} G_F\over 16\pi^2}
\hbox{Im}(\lambda_{tt}  \lambda_{bb})
{2\over3} H\left(m_t^2\over M^2_{H^\pm}\right)
\left( {\alpha_s(m_W)\over \alpha_s(m_b)} \right)^{14\over 23}
\ ,
\end{equation}
where the function $H$ is
\begin{equation}
H(y)={3\over2}{y\over (1-y)^2} \left(y-3-{2\log y\over 1-y}\right) \ .
\end{equation}
Note that $H(1)=1$ when $M_{H^\pm}=m_t$.  Numerically,
\begin{equation}
d^g_n=10^{-25}\hbox{e$\cdot$cm}
\  \hbox{Im}(\lambda_{tt}  \lambda_{bb})
\left({\alpha(m_n)\over\alpha(\mu)}\right)^{1\over2}
\left({\xi_g\over 0.1}\right)
H\left({m_t^2\over M^2_{H^\pm}}\right) \ .
\end{equation}
The experimental limit,
\begin{equation}
d_n<10^{-25}\hbox{e$\cdot$cm} \ ,
\end{equation}
places an upper bound
$|\hbox{Im}(\lambda_{tt}\lambda_{bb}  )| \alt 1 $
on the coupling product for our choice of parameters, $\xi_g=0.1$,
$\mu=m_n$ when $M_{H^\pm}\simeq m_t$.
The bound is sensitive to  uncertainties in  $\mu$ and
$\xi_g$, but not much in  $M_{H^\pm}$. The function value $H$ decreases
only by a factor of 1.6 as  the charged Higgs mass varies from 50 GeV to
200 GeV.

In Fig.~1, the constraint on the $M_{H^\pm}$ versus
$\theta$ is given by the shaded areas which are excluded by the
NEDM measurement.

For the case of rather large $|\lambda_{tt}\lambda_{bb}|\gg 1$, the
phase becomes restricted to the forward region $\theta \sim 0$ or the
backward region $\pi$. However, the backward region ($\theta\sim\pi$) is not
preferable for  $M_{H^\pm} \alt 500$ GeV due to the constraint from
$b\rightarrow s\gamma$.
If the charged Higgs boson is this light with large couplings to the
$b$ and $t$ quarks, the NEDM analysis requires a small phase in the forward
region.
On the other hand, when $|\lambda_{tt}\lambda_{bb}| < 0.7$, the NEDM
constraint becomes ineffective and the constraint from $b\to s\gamma$
remains useful.

Other places to look for the effects of this angle $\theta$ include other
$b\to s,d$ decays, CP violation effects in $b\to s\gamma$ \cite{Wolfenstein},
$b\to s \ell \bar \ell$,
and  the electric dipole moments of fermions via a 2-loop mechanism
\cite{dave}.

On the other hand,
this phase angle $\theta$ will not show up in other existing
constraints like $\rho_0$, $R_b$, and flavor-mixing.  The previous
argument that the 2HDM only has a very narrow window left to accommodate all
the constraints from $B(b\to s\gamma)$, $\rho_0$, $R_b$, and flavor-mixing
is now not true because of the possible phase angle in model III that we are
considering.  The narrow window on $M_{H^\pm}$ opens up.
We shall summarize the other constraints on $\lambda_{tt}$,
$\lambda_{bb}$, and Higgs masses in the next section.

\section{Other Constraints}

Direct searches for Higgs bosons in 2HDM at LEPII \cite{ichep98} place
the following limits on Higgs boson masses:
\begin{equation}
\label{direct}
M_{h^0} > 77 \;{\rm GeV}\;, \qquad M_A> 78 \;{\rm GeV}\;, \qquad
M_{H^\pm} > 56-59\;{\rm GeV} \;,
\end{equation}
where the $M_{h^0}$ and $M_A$ mass limits are obtained by combining
the four LEP experiments but no combined limit on $M_{H^\pm}$ is available
\cite{ichep98}.  We shall then discuss other constraints from
precision measurements.

\subsection{$K^0-\overline{K^0}$, $D^0-\overline{D^0}$,
and $B^0-\overline{B^0}$}

These $F^0-\overline{F^0}$ ($F=K,D,B$) flavor-mixing processes can
occur via tree-level, penguin, and box diagrams in model III
\cite{atwood}.  One particular argument against the model III is that
it allows FCNC at the tree-level, but with a lot of freedom in picking the
parameters $\lambda_{ij}$ it certainly survives all the present FCNC
constraints.  The tree-level diagrams for these $\Delta F=2$ processes
can be eliminated by choosing $\lambda_{ui},\lambda_{dj}$ very small.
Actually, in our study we have set $\lambda_{ij}=0\; (i\neq
j)$, therefore, all tree-level FCNC diagrams are eliminated and so do the
penguin diagrams.  However, there are important contributions coming
from the box diagrams with the charged Higgs boson.  Naively, to
suppress the charged Higgs contribution we need to increase the
charged Higgs mass or decrease $\lambda_{tt}$.  We shall obtain a set
of bounds using the experimental measurement $x_d$ of
$B^0-\overline{B^0}$ in the following (
$K^0-\overline{K^0}$ and $D^0-\overline{D^0}$ mixings are small in our model
because  of the mass hierarchy choice of $\hat \xi^{U,D}_{ij}$ in Eq.
(\ref{anat})).

\begin{figure}[ht]
\leavevmode
\begin{center}
\includegraphics[height=5in]{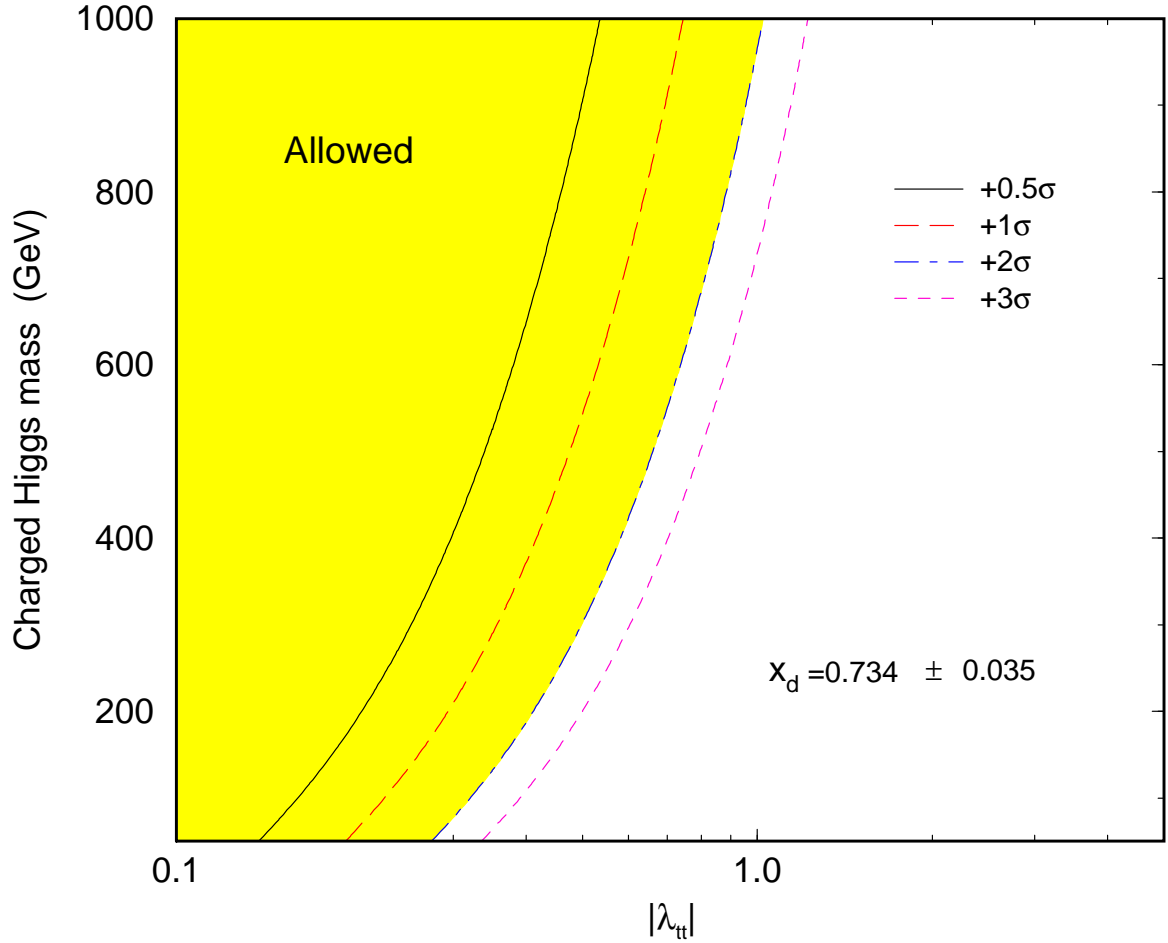}
\end{center}
\caption{Contour plot of the $B^0-\overline{B^0}$ mixing parameter $x_d$
in the plane of $|\lambda_{tt}|$ and the charged Higgs boson mass.
The experimental value is $x_d=0.734\pm0.035$.}
\label{xb}
\end{figure}

The quantity that parameterizes the $B^0-\overline{B^0}$ mixing is
\begin{equation}
x_d \equiv \frac{\Delta m_B}{\Gamma_B} = \frac{G_F^2}{6\pi^2} |V_{td}^*|^2
|V_{tb}|^2 f_B^2 B_B m_B
\eta_B \tau_B M_W^2 \left(I_{WW} + I_{WH}+I_{HH} \right)
\end{equation}
where\cite{abbott}
\begin{eqnarray}
I_{WW} &=& \frac{x}{4} \biggr[ 1 + \frac{3-9x}{(x-1)^2} +
 \frac{6x^2 \log x}{(x-1)^3} \biggr] \nonumber \\
I_{WH} &=& xy |\lambda_{tt}|^2 \biggr[ \frac{(4z-1)\log y}{2(1-y)^2(1-z)}
 -\frac{3\log x}{2(1-x)^2(1-z)} +\frac{x-4}{2(1-x)(1-y)} \biggr] \nonumber \\
I_{HH} &=& \frac{xy |\lambda_{tt}|^4}{4} \biggr[ \frac{1+y}{(1-y)^2} +
  \frac{2y\log y}{(1-y)^3} \biggr]\nonumber  \;,
\end{eqnarray}
where $x=m_t^2/M_W^2$, $y=m_t^2/M_{H^\pm}^2$, $z=M_W^2/M_{H^\pm}^2$, and
the running top mass $m_t=m_t(m_t)=166$ GeV.
We use these inputs \cite{pdg,langacker,buras}: $|V_{tb}|=1$, $f_B^2 B_B=
(0.175\;{\rm GeV})^2(1.4)$, $m_B=5.2798$ GeV, $\eta_B=0.55$, and
$x_d=0.734\pm 0.035$, $\tau_B=1.56$ ps.
Since the allowable range of $|V_{td}|$ is from 0.004 to 0.013
\cite{pdg}, we use a central value for $|V_{td}|$ obtained using the
central value of $x_d$ and it gives $|V_{td}| \simeq 0.0084 $ 
(which is the central value given in the Particle Data Book 98 \cite{pdg}.)
We then obtain bounds on $\lambda_{tt}$ and $M_{H^\pm}$ by the $2\sigma$
limit of $x_d$ assuming the only error comes from $x_d$ measurement
(see Fig.~\ref{xb}):
\begin{equation}
M_{H^\pm} \agt 77\; (60) \ {\rm GeV} \qquad
{\rm for} \qquad |\lambda_{tt}| \alt 0.3\;(0.28) \;,
\end{equation}
which means virtually no limit on the charged Higgs mass if
$|\lambda_{tt}| \alt 0.28$, because the present
direct search limit on charged Higgs boson is about 56--59 GeV
(Eq.(\ref{direct})).
We have improved the results in Ref. \cite{grant,atwood}
because we are using an updated value of $x_d$.  In the context of model
I and II the bound is $M_{H^\pm}\agt 77(60)$ GeV for $\tan\beta \agt 3.3(3.6)$.
For $\tan\beta$ gets close to 1, $M_{H^\pm}>1$ TeV.

\subsection{$\rho_0$}

$\rho$ was introduced to measure the relation between the masses of $W$ and
$Z$ bosons.  In the SM $\rho=M_W^2/M_Z^2 \cos\theta_{\rm w}=1$ at the
 tree-level.
However, the $\rho$ parameter receives contributions from the
SM corrections and from new physics.  The deviation from the SM predictions
is usually described by the parameter $\rho_0$ defined by \cite{langacker}
\begin{equation}
\rho_0 = \frac{M_W^2}{\rho M_Z^2 \cos^2 \theta_{\rm w}} \,,
\end{equation}
where the $\rho$ in the denominator absorbs all the SM corrections, among which
the most important SM correction at 1-loop level comes from the heavy
top-quark:
\begin{equation}
\rho \simeq 1+ \Delta\rho_{\rm top} =
1 +  \frac{3G_F}{8 \sqrt{2} \pi^2} \, m_t^2\,,
\end{equation}
in which $\Delta \rho_{\rm top}$ is about 0.0095 for $m_t=173.8$ GeV.  By
definition  $\rho_0 \equiv 1$ in the SM.
The reported value of $\rho_0$ is \cite{langacker}
\begin{equation}
\label{rho0}
\rho_0 = 0.9996 \stackrel{\scriptstyle +0.0017}{\scriptstyle -0.0013}
\;\; (2\sigma)\,.
\end{equation}
In terms of new physics (2HDM here) the constraint becomes:
\begin{equation}
-0.0017 < \Delta \rho_{\rm 2HDM} < 0.0013 \;.
\end{equation}
In 2HDM $\rho_0$ receives contribution from the Higgs bosons given by,
in the context of model III,  \cite{grant,denner,atwood}
\begin{equation}
\label{rho}
\Delta \rho_{\rm 2HDM} = \frac{G_F}{8\sqrt{2}\pi^2} \biggr [
\sin^2 \alpha F(M_{H^\pm}, M_A, M_{H^0}) +\cos^2\alpha F(M_{H^\pm}, M_A,
 M_{h^0})
\biggr] \,,
\end{equation}
where
\begin{eqnarray}
F(m_1,m_2,m_3) &=& m_1^2 -\frac{m_1^2 m_2^2}{m_1^2 - m_2^2}
 \log\left( \frac{m_1^2}{m_2^2} \right) \nonumber \\
&& -\frac{m_1^2 m_3^2}{m_1^2 - m_3^2}
 \log\left( \frac{m_1^2}{m_3^2} \right)
+\frac{m_2^2 m_3^2}{m_2^2 - m_3^2}
 \log\left( \frac{m_3^2}{m_3^2} \right) \nonumber \;.
\end{eqnarray}
Since $\rho_0$ is constrained to be around 1 we have to minimize the
contributions of $\Delta\rho_{\rm 2HDM}$.
Without loss of generosity we set $\alpha=0$, which means that the heavier
neutral Higgs $H^0$ decouples and the first Higgs doublet can be identified
as the SM Higgs doublet, while the second Higgs doublet is the source of
new physics.  The leading behavior of $\Delta\rho_{\rm 2HDM}$ scale as
$M_{H^\pm}^2$ and, therefore, the constraint of $\rho_0$ in Eq. (\ref{rho0})
puts an upper bound on $M_{H^\pm}$.
Actually,  if the charged Higgs mass $M_{H^\pm}$ is between $M_A$ and $M_{h^0}$
the $\Delta\rho_{\rm 2HDM}$ is negative.  However, this is not the favorite
scenario because in the case of $R_b$ the experimental result prefers
$M_A\simeq M_{h^0} \approx 80-120$ GeV,
that will be discussed in the next subsection.
In this case $M_A\simeq M_{h^0}$, $\Delta\rho_{\rm 2HDM}$ is positive and,
therefore, we want to keep it small.  Using Eq. (\ref{rho}) for
$M_A\simeq M_{h^0} = 80 - 120$ GeV, the charged Higgs mass is constrained to be
\begin{equation}
M_{H^\pm} \alt 180 - 220  \;{\rm GeV} \;.
\end{equation}

\subsection{$R_b$}

$R_b$ was about $+3.7\sigma$ above the SM value a few years ago, but now the
deviation is reduced to $+1\sigma$ after almost all LEP data have been
analyzed \cite{langacker}.  $R_b^{\rm exp}$ still places a constraint on
the 2HDM, though it is much less severe than before.
This is because only a narrow window exists in the
neutral Higgs bosons that does not decrease $R_b$ while the charged-Higgs boson
always decreases $R_b$.  We shall divide the discussion into two parts:
neutral-Higgs contribution and charged-Higgs contribution.

\begin{figure}[ht]
\leavevmode
\begin{center}
\includegraphics[width=4.5in]{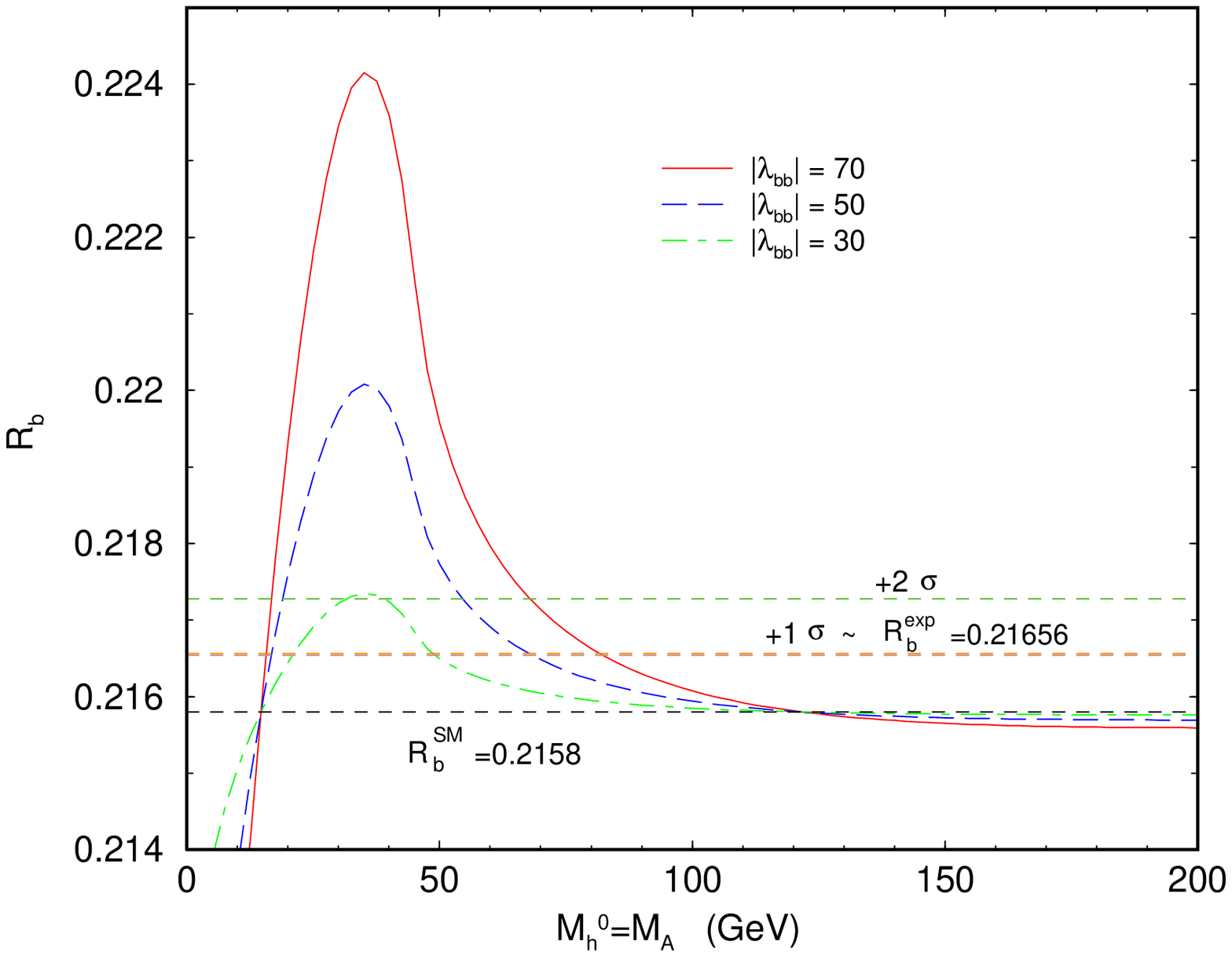}
\includegraphics[width=4.5in]{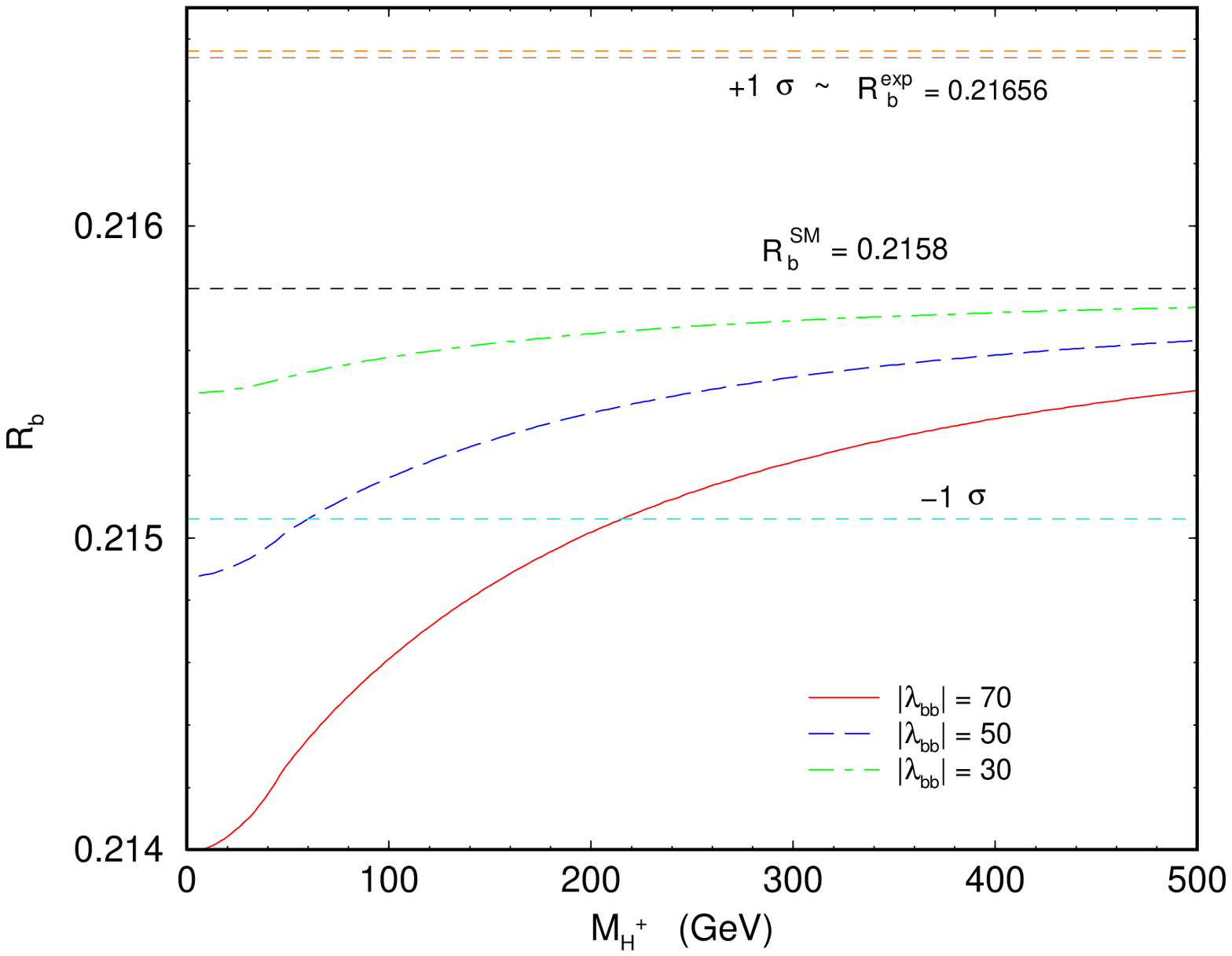}
\end{center}
\caption{The $R_b \equiv
\Gamma(Z\to b\bar b)/\Gamma_{\rm had}$ due to the presence
of (a) the neutral Higgs bosons, $M_A\simeq M_{h^0}$ and (b) the charged Higgs
boson. }
\label{rb}
\end{figure}

According to Ref.~\cite{denner} the contribution from the neutral Higgs
boson is positive in a narrow window of $20\;{\rm GeV} < M_A \simeq M_{h^0}
< 120$ GeV and is negative otherwise.
Since the charged Higgs boson contribution  always decreases $R_b$, it
makes more sense to require  the neutral Higgs contribution to be positive.
Here we adapt the formulas in Ref.~\cite{denner}
to model III.  First, the contribution from the neutral Higgs bosons only
depends on $|\lambda_{bb}|$ and the masses of the neutral Higgs bosons.
Again without loss of generosity, we set the scalar Higgs-boson mixing
angle $\alpha=0$  in order to decouple the heavier $H^0$.
We show the resultant $R_b$ due to the presence of the neutral Higgs bosons
in Fig. \ref{rb}(a) for $|\lambda_{bb}|=30,50,70$,
where $R_b^{\rm SM}=0.2158$, $R_b^{\rm exp}=0.21656\pm0.00074$
\cite{langacker},
 and the $1\sigma$ is taken to be the standard deviation of the
experimental result.  In Fig. \ref{rb}(a) the horizontal lines
represent the $R_b^{\rm SM}$, $+1\sigma$, and $+2\sigma$ values.
The $R_b^{\rm exp}$ is almost at the $+1\sigma$ line.
If we allow only $1\sigma$ value below $R_b^{\rm exp}$, we need
$M_{h^0} \approx M_A \approx 80-120$ GeV with a fairly large $|\lambda_{bb}|$.
For $|\lambda_{bb}|$ as large as 70 the enhancement can be as large as
$+1\sigma$ at $M_{H^\pm}=80$ GeV.
On the other hand, if we allow $2\sigma$ below $R_b^{\rm exp}$, then we
can have all the range of $M_{h^0}\approx M_A >80$ GeV, as can be seen in
Fig. \ref{rb}(a).
At any rate, the preferred scenario is $M_{h^0}\approx M_A=80-120$ GeV with
a fairly large $|\lambda_{bb}|$.  How large $|\lambda_{bb}|$ should be?
It depends on the charged Higgs contribution as well.

Since the charged Higgs-boson contribution is always negative,
we want to make it as small as possible.  This contribution depends
on $|\lambda_{tt}|$, $|\lambda_{bb}|$, and $M_{H^\pm}$.
The effect on $R_b$ due to the presence of the charged Higgs boson is shown in
Fig. \ref{rb}(b) for $|\lambda_{bb}|=30, 50,70$ and $|\lambda_{tt}|=0.05$.
In Fig. \ref{rb}(b) the horizontal lines represent the
$R_b^{\rm SM}$ and $\pm 1\sigma$.
The $R_b^{\rm exp}$ is very close to $+1\sigma$ line.
It is clear from the graph that because
we do not want the charged Higgs contribution
to reduce $R_b^{\rm SM}$ by more than $1\sigma$, we require
$M_{H^\pm} \agt 60\;(220)$ GeV for $|\lambda_{bb}|=50\;(70)$.

Since $R_b^{\rm exp}$ is only $+1\sigma$ away from $R_b^{\rm SM}$,
it is not necessary to keep the narrow window of $M_A$ and $M_{h^0}$ if
we allow $2\sigma$ below the experimental data.  In this case,
$M_{h^0}$ and $M_A$ can be widened to much larger masses, and so the
$\rho_0$ constraint on the ceiling of the charged Higgs mass will also
be relaxed.
However, $M_{H^\pm}$ cannot be too small otherwise $R_b$ will be decreased to
an unacceptable value.

Summarizing this section the constraints by $B^0-\overline{B^0}$ mixing,
$\rho_0$, and $R_b$ give the following preferred scenario:
\begin{enumerate}
\item $M_A \simeq M_{h^0} = 80-120$ GeV;
\item $|\lambda_{bb}| \simeq 50$;
\item $|\lambda_{tt}| \alt 0.3$;
\item $80$ GeV $\alt M_{H^\pm} \alt 200$ GeV.
\end{enumerate}

\section{Conclusions}

We have demonstrated in model III of the general two-Higgs-doublet model
the charged-Higgs-fermion couplings can be complex, even in the simplified
case of no tree-level FCNC couplings.  The phase angle in the complex
charged-Higgs-fermion coupling determines the interference between the
standard model amplitude and the charged-Higgs amplitude in the process
of $b\to s\gamma$.  We found that for $|\lambda_{bb} \lambda_{tt}| \simeq 1$
there is a large range of the phase angle ($\theta \approx 50^\circ-90^\circ$
and $270^\circ-310^\circ$)
such that the rate of $b\to s\gamma$ is within the experimental
value for all range of $M_{H^\pm}$.  In other words, the previous tight
constraints on $M_{H^\pm}$ from the CLEO $b\to s\gamma$ rate is relaxed,
depending on this phase angle.  In addition, we also examined the effect
of this phase angle on the neutron electric dipole moment and discussed
other experimental constraints on model III.  The necessary constraints
are already listed at the end of the last section.
Here we offer the following comments:

\begin{enumerate}

\item The phase angle induces a CP-violating chromoelectric dipole moment
of the $b$-quark, which leads to a substantial enhancement in
 neutron electric dipole moment.
The experimental upper limit on neutron electric dipole moment thus places
a upper bound on the couplings: $|\lambda_{tt}\lambda_{bb}| \sin\theta \alt 
0.8$ for $M_{H^\pm}\approx 100$ GeV.
This bound has large uncertainties due to the hadronic matrix element of
the neutron and the factorization scale.

\item The phase angle will also cause other CP-violating effects in other
processes, e.g., the decay rate difference between $b\to s\gamma$ and
$\bar b \to \bar s \gamma$ \cite{Wolfenstein}, and in lepton asymmetries of
$b \to s \ell^+ \ell^-$.  These	 processes will soon be measured at the
future $B$ factories.

\item Other experimental measurements, like $F^0-\overline{F^0}$ mixing,
$\rho_0$, and $R_b$, constrain only the magnitude of the couplings and the
Higgs-boson masses but not the phase angle.

\item The $B^0-\overline{B^0}$ mixing measurement can only constrain the
charged-Higgs mass and $|\lambda_{tt}|$ loosely because the mixing
parameter $x_d$ depends on $|V_{td}|$, which is not yet well measured.
Other uncertainties come from the hadronic factors: $f_B$, $B_B$, and
$\eta_B$.
Actually, the mixing parameter $x_d$ is often used to determine $|V_{td}|$.

\item As we have mentioned, if $R_b^{\rm SM}$ gets closer to
the SM value the constraint on the neutral Higgs-boson masses:
$M_A$ and $M_{h^0}=80-120$ GeV will go away completely.  On the other
hand, the charged-Higgs boson mass is still required to be larger than
about 60 GeV (for $|\lambda_{bb}|=50$)
in order not to decrease $R_b$ significantly.

\end{enumerate}

\section*{\bf Acknowledgments}
This research was supported in part by the U.S.~Department of Energy under
Grants Nos. DE-FG03-93ER40757, DE-FG02-84ER40173, and DE-FG03-91ER40674 and
by the Davis Institute for High Energy Physics.


\end{document}